\documentclass[
aps,nofootinbib,showpacs,showkeys,preprint
tightenlines,preprintnumbers,] {revtex4}

\usepackage{epsf,epsfig,subfigure,graphicx,amsmath,amssymb}
\usepackage{color}
\newcommand{\dis}[1]{\begin{equation}\begin{split}#1\end{split}\end{equation}}

\newcommand{\gev}{\,\textrm{GeV}}

\newcommand{\eV}{\,\mathrm{eV}}
\newcommand{\SMSSM}{$\sigma$MSSM}

\newcommand{\Mp}{{M_{\rm P}}}
\newcommand{\Mg}{{M_{\rm GUT}}}
\newcommand{\ie}{{\it i.e.}\ }
\newcommand{\etal}{{\it et al.}}
\newcommand{\Z}{{\bf Z}}
\newcommand{\NDW}{{N}_{\rm DW}}

\def\E6{{\rm E_6}}
\def\EE8{{\rm E_8\times E_8}}

\begin{document}

\title{Axionic domain wall number related to U(1)$_{\rm anom}$ global symmetry}

\author{Jihn E.  Kim}
\address
{Center for Axion and Precision Physics Research (IBS),
  291 Daehakro,  Daejeon 34141, Republic of Korea, and\\
Department of Physics, Kyung Hee University,\\
 26 Gyungheedaero,  Seoul 02447, Republic of Korea, and\\
Department of Physics, Seoul National University, 
 1 Gwanakro,  Seoul 08826, Republic of Korea 
}
\begin{abstract}
The QCD axion with $f_a$ at an intermediate scale, $ 10^{9\,}\gev\sim10^{12\,}\gev $, seems in conflict with the gravity spoil of global symmetries and may face the axionic domain wall problem. We point out that the string compactifications with an anomalous U(1) gauge symmetry, allowing desirable chiral matter spectra, circumvent these two problems simultaneously. 

\end{abstract}


\maketitle

\section{Introduction}

At present, it is a challenging issue to detect any cosmological hint of bosonic collective motion (BCM)  \citep{kim14}. The QCD axion is the most studied pseudoscalar boson in this context. Firstly, it is designed to solve the strong CP problem under the Peccei-Quinn (PQ) symmetry \citep{PQ77, WW78}. But, only the weak interaction singlets are useful for this type of the strong CP solution \citep{KSVZ1,KSVZ2,DFSZ}. Then, for such `invisible axions', their astrophysical effects  have led to the following three important constraints on the axion parameters: from the BCM energy density \citep{BCMorig83}, axionic string and domain wall networks \citep{SikivieDW}, and the energy loss mechanism in big stars \citep{KimPRP87,SNstudy88},
\dis{
10^{9\,}\gev\lesssim f_a\lesssim  10^{12\,}\gev,~~\NDW=1, \label{eq:Consts}
}
where $f_a$ is the axion decay constant and $\NDW$ is the domain wall number in the axion model.

From the theoretical perspective of fulfilling beyond-the-standard-model (BSM) completion of the Standard Model (SM), the role of the SM singlets seems inevitable \citep{KSVZ1}. The simplest BSM is the grand unification (GUT) SU(5), which needs at least one SM singlet in the Higgs field {\bf 24}. In fact, that singlet was used for a GUT scale axion \citep{GGW81}. More importantly,  ultra-violet completions of the SM have been tried in string compactification, where   color and weak interaction singlets are numerous in general. 

In string theory, in addition to singlets from matter fields one must deal with the SM singlet fields from the 10 dimensional (10D) antisymmetric tensor gauge field  $B_{MN}$, where $M,N\in\{1,2,\cdots, 10\}$.  The so-called Green-Schwaz (GS) mechanism
requires this to be realized toward a consistent anomaly-free 10D field theory \citep{GS84}.
Compactifying the 10D string theory down to 4 dimensional (4D) Minkowski space,  $B_{MN}$ splits into numerous singlets: to   $B_{\mu\nu}$ with the Minkowski indices $\mu,\nu\in\{1,2,3, 4\}$ and   $B_{ij}$ with the internal indices $i,j\in\{5,6,\cdots, 10\}$. $B_{\mu\nu}$ is the so-called model-independent (MI) axion \citep{WittenMI} and $B_{ij}$ is the  model-dependent (MD) axion \citep{WittenMD}. The relevant high energy scales in string compactification are the string scale $M_s\approx 0.7\times 10^{18\,}\gev$, the GUT scale  $\Mg\simeq  2\times 10^{16\,}\gev$, and the gaugino condensation scale $M_{\Lambda_G}\approx   10^{13\,}\gev$ or supersymmetry(SUSY) breaking scale  $F_S\simeq M_{\Lambda_G}^3/\Mp$ \citep{Nilles84}. Among these, the MI axion has a rather well-defined scale around $f_a\approx 10^{15\,}\gev$ \citep{ChoiKim85}, but it is outside the constraint (\ref{eq:Consts}).

The upper bound on $f_a\approx 10^{12\,}\gev$ in (\ref{eq:Consts})
should be clarified. Firstly, it is the value obtained from the assumption that the Universe is closed by the QCD axion plus baryons. So, if there are additional components of dark matter, the upper bound should be smaller than that. Second, it is obtained from the coefficient $\frac{\langle a\rangle}{f_a}$ of the gluon anomaly term  $\{G^\alpha \tilde{G}^\alpha\}\equiv (1/32\pi^2)G^{\alpha}_{\mu\nu}\tilde{G }^{\alpha\,\mu\nu}$. But the value $f_a$ itself is not the vacuum expectation value (VEV), $ \langle \sigma\rangle$, of the SM singlet field $\sigma$, breaking the PQ symmetry. They are related by the domain wall number $\NDW$: $\langle \sigma\rangle=\NDW f_a/\sqrt2$ \citep{SikivieDW}. If $\NDW$ is large, $\langle \sigma\rangle$ can be closer to the GUT scale. Indeed, the QCD axion from string compactification can give the values $\sqrt2 \langle \sigma\rangle\gg f_a$ \citep{KimPLB88,KimPLB14}. Third, the physical domain wall number is determined from the first guess  by modding out by the number of vacuum degeneracy \citep{LS82,ChoiKimDW85}.
In this paper, we discuss the domain wall number related to the MI axion $B_{\mu\nu}$ and the spontaneously broken global symmetry U(1)$_{\rm anom}$.

\section{Domain wall number constraint}
\label{sec:2}

\begin{figure}[!t]
\begin{center}
\includegraphics[width=0.60\linewidth]{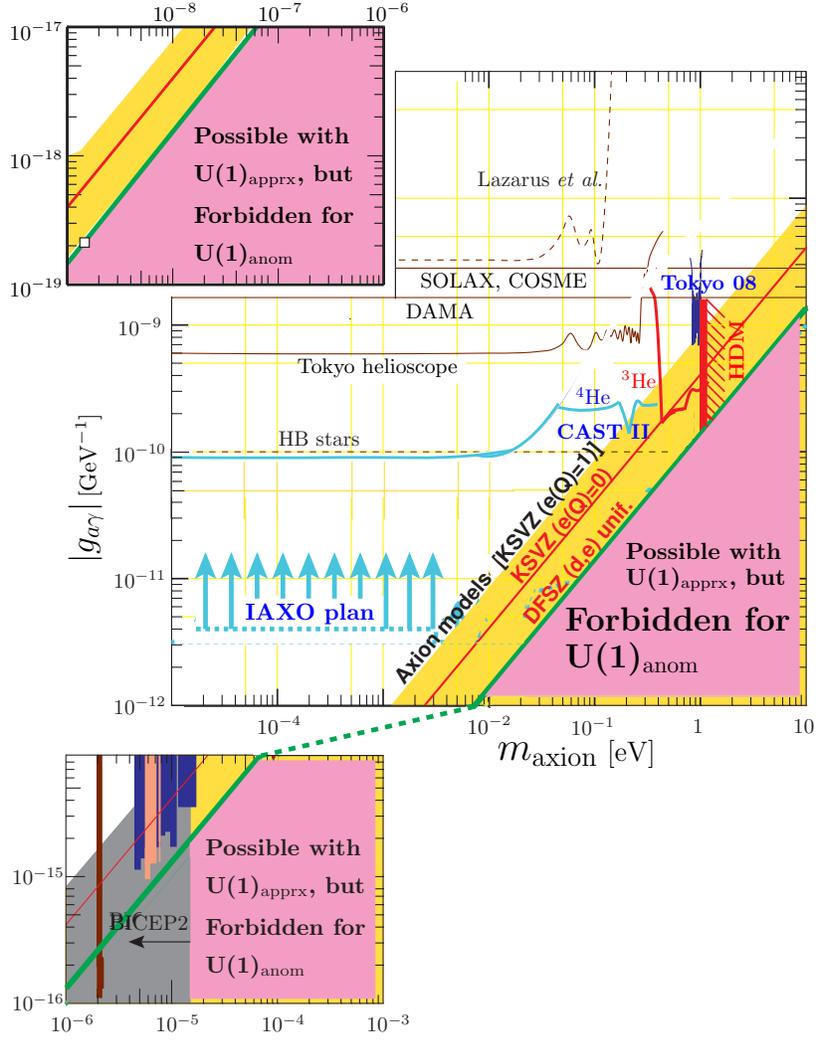}
\end{center}
\caption{The axion search plot in the plane of $g_{a\gamma}\, [\gev^{-1}]= 1.57\cdot 10^{-10}\,c_{a\gamma\gamma}$ vs. $m_a$. Model lines are from \citep{KimRMP10,KimPRD98}. The MI axion point, in case there is no gauge anomaly below the compactification scale, is shown as the white square (Ref. \citep{ChoiKim85}) in the upper left corner.  } \label{fig:Uanom}
\end{figure}

The solution of the strong CP problem by the intermediate scale $f_a$ of (\ref{eq:Consts}), including the gravity effects, requires that many low order terms in the (super)potential must be forbidden  \citep{BarrGr92}, which is a kind of fine-tuning. In this regard, we point out that there arise numerous `approximate' U(1) global symmetries in string compactification   \citep{KimPLB13}, among which those cancelling many low order terms are possible candidates for the  PQ symmetry toward the QCD axion \citep{ChoiKimIW07}.  However, a still better candidate is an exact global   symmetry for the QCD axion. This possibility is provided by the MI axion $B_{\mu\nu}$ \citep{GS84,WittenMI}.  $B_{\mu\nu}$ behaves like a gauge field under local transformation, and hence it may be free from the gravity obstruction of global symmetries. Because only transverse fields are physical gauge fields, the antisymmetric tensor $B_{\mu\nu}$ contains physical massless components with the number of  degrees $ (n-2)!/2! =1$ in 4D. It is most easily seen by the field strength $H_{\mu\nu\rho}\equiv  \partial_{[\mu}\,B_{\nu\rho]}\equiv f_a'\epsilon_{\mu\nu\rho\sigma}\partial^\mu a_{MI}$ with $f_a'= \Mp/6\sqrt2 $ where the above one transverse degree is expressed as the MI axion with the duality transformation. The original kinetic energy term of $B_{\mu\nu}$ becomes the kinetic energy term of $a_{MI}$ \citep{ChoiKim85},
\dis{
\frac{3k^2}{2g^4\phi^2}\, H_{\mu\nu\rho}H^{\mu\nu\rho}\to \frac12\,\partial^2 a_{MI},
}
with \cite{KimPRP87}
\dis{
f_{a_{MI}}=\frac{f_a'}{8\pi^2}= \frac{\Mp}{48\sqrt2 \pi^2}\simeq 3.63\times 10^{15\,}\gev.\label{eq:fvsfp}
}
The Green-Schwarz mechanism gives the 4D equation of  $a_{MI}$ \cite{WittenMI},
\dis{
&\partial^2 a_{MI} =-\frac{1}{32\pi^2\,f_{a_{MI}} }\left(  G^a\tilde{G}^a+
  W^i\tilde{W}^i+\cdots \right),\\
&G^{\alpha}\tilde{G}^{\alpha} = \frac{1}{2} \epsilon^{\mu\nu\rho\sigma}G^{\alpha}_{\mu\nu} G^{\alpha}_{\rho\sigma} ,~W^i\tilde{W}^i = \frac{1}{2} \epsilon^{\mu\nu\rho\sigma}W^i_{\mu\nu} W^i_{\rho\sigma} ,~\cdots \label{eq:AnomCoupl}
}
where $G^{\alpha}_{\mu\nu}\,({\alpha}=1,2,\cdots,8)$ and $W^i_{\mu\nu}\,(i=1,2,3)$ are the field strengths of gluon and SU(2)$_L$ fields in the SM, and $\cdots$ denotes other possible non-Abelian gauge fields and the U(1) fields. The GS mechanism gives the equation of motion of   $a_{MI}$ with exactly the same coefficient for all gauge (non-Abelian and properly normalized Abelian) anomalies as implied in Eq. (\ref{eq:AnomCoupl}). The axion-photon-photon coupling follows the line of unification point because the MI axion couples to all gauge anomalies universally as shown in Eq. (\ref{eq:AnomCoupl}) \citep{KimAxCoupl16}.  The value $f_{a_{MI}}$ was calculated at the order $\approx 3.6\times 10^{15\,}\gev$ which is marked as the white square in the axion search plot of Fig. \ref{fig:Uanom}.

From Eq. (\ref{eq:AnomCoupl}), one can consider the following effective interaction of the MI axion with gluon fields, 
\dis{
{\cal L}_{a_{MI}}=\frac12 \partial^\mu a_{MI}\partial_\mu a_{MI} -\,\frac{  a_{MI}}{ f_{a_{MI}} } \left(\frac{  1 }{32\pi^2 } G^{\alpha}_{\mu\nu}\tilde{G}^{\alpha,\mu\nu}\right)
}
which defines in fact the axion decay constant such that the coefficient of the anomaly term is $a/f_a$ \citep{KimRMP10}. The action $S$ due to the anomaly term is basically the Pontryagin index which is $\pm 1$ for the instanton solution of Belavin \etal\,\citep{Belavin74}. Therefore, since the shift of $a_{MI}$ to  $a_{MI}+2\pi f_{a_{MI}}$ returns $e^{iS}$ to its original value, the MI-axion vacuum returns to itself. The question is how the other matter fields transform under this shift of $a_{MI}$. Since there is no matter coupling of $a_{MI}$, the periodicity $2\pi f_{a_{MI}}$ is the periodicity in the full compactified theory. Thus, the domain wall number of $a_{MI}$ is one, which was shown in Ref. \citep{WittenMD}. Here, we present another proof that the domain wall number of $a_{MI}$ is one.

A horizon scale cosmological string is created by the Kibble mechanism when a global U(1) symmetry is spontaneously broken \citep{Kibble80prp}. What can be the corresponding source of string for the MI axion? The MI-axion does not have the string configuration by the Kibble mechanism, but still the axionic string must result below the compactification scale. The pure gauge configuration is the origin of MI-axionic string. It is the pure gauge function $\Lambda_\mu(x)$ making $H_{\mu\nu\rho}$ gauge invariant under the gauge transformation.
By mapping the axion field $a_{MI}$ to the azimuthal angle $\theta$ (in the cylindrical polar coordinate) around the string, if the domain wall number of $a_{MI}$ is $\NDW(a_{MI})$  then the action Eq. (\ref{eq:AnomCoupl}) returns to itself after the shift of $a_{MI}$ by $a_{MI}\to a_{MI}+2\pi\,\NDW(a_{MI}) f_{a_{MI}}$, \ie the gluon coupling must have the form
\dis{
-\,\frac{  a_{MI}}{  \NDW(a_{MI})f_{a_{MI}} } \left(\frac{1 }{32\pi^2} G^{\alpha}_{\mu\nu}\tilde{G}^{\alpha,\mu\nu}\right),\label{eq:LarDWN}
}
since the smallest Pontryagin index is defined to be $\pm 1$,
\dis{
\int d^4x\left(\frac{1 }{32\pi^2} G^{\alpha}_{\mu\nu}\tilde{G}^{\alpha,\mu\nu}\right)=\pm 1.\label{eq:Pontryagin}
}
In other words, if $\NDW(a_{MI})>1$, the action (\ref{eq:LarDWN}) does not return to itself for the shift  $a_{MI}\to a_{MI}+2\pi\,n f_{a_{MI}}$ for $n=\{1,2,\cdots,\NDW(a_{MI})-1\}$. Thus, (\ref{eq:LarDWN}) describes the MI axion coupling if its domain wall number is $\NDW(a_{MI})$. In this case, integration of Eq. (\ref{eq:LarDWN}) is

\dis{
 \frac{1}{32\pi^2}\int d^4x\, G^\alpha_{\mu\nu}\tilde{G}^{\alpha,\mu\nu} &=  \NDW(a_{MI})f_{a_{MI}}\int d^4x\,   \partial^\mu\partial_\mu a 
=\NDW(a_{MI})f_{a_{MI}} \int d^4x\,\partial_\alpha \frac{1}{3! f_a'}\epsilon^{\alpha\nu\rho\sigma}
\partial_{[\sigma}\,B_{\nu\rho]} \\
&= \frac{\NDW(a_{MI})}{3! 8\pi^2} \int d^4x\,\partial_\alpha   \epsilon^{\alpha\nu\rho\sigma}
\partial_{[\sigma}\,B_{\nu\rho]},\label{eq:Int1}
}
where we used the relation (\ref{eq:fvsfp}).
Let us perform the 4D integral (\ref{eq:Int1}) in the direct product space $dz\times d\Omega_z$ where $\Omega_z$ is the 3D surface orthogonal to $z$.\footnote{Even though the heterotic string is  closed, we can consider a large cosmological string. Putting a point of the large string at origin, we can consider an almost straight line along the $z$ axis.} The 4D Gauss theorem gives
\dis{
 \int d^4x\, \partial^\alpha  \Omega_\alpha    =\int d^4x\, \partial^z \Omega_z=\int d^3x\,   \Omega_z ,\label{eq:Int3}
}
where in the Euclidian space, $d^3x=rd\theta dr dt=\frac12 d\theta dt dr^2$,
\dis{
\Omega_z =\epsilon_{z\nu\rho\sigma}
\partial^{[\sigma}\,B^{\nu\rho]}.
}

The gauge function symmetric around $z$-axis of 3D cylindrical polar coordinate, $(r,\theta,z)$, is
\dis{
&\Lambda_{\theta } =\frac{\xi a^3}{(r^2+ t^2+a^2)^2} ,\, \Lambda_{r\theta } =\frac{-4\xi a^3 r}{(r^2+ t^2+a^2)^3} ,\,\Lambda_{t\theta } =\frac{-4\xi a^3 t}{(r^2+ t^2+a^2)^3}, \\
& \partial_t \Lambda_{r\theta} =\frac{24\xi a^3 rt}{(r^2+ t^2+a^2)^4} ,\, \partial_r \Lambda_{t\theta} =\frac{24\xi a^3 rt}{(r^2+ t^2+a^2)^4} ,
}
\dis{
&\epsilon^{z\mu\nu\rho} \partial_{[\mu}\Lambda_{\nu\rho]} =2\epsilon^{z trz} \frac{24\xi a^3 rt}{(r^2+ t^2+a^2)^4}
=  \frac{48\xi a^3 rt}{(r^2+ t^2+a^2)^4}.
\label{eq:IntGauge}
}
Integration of (\ref{eq:IntGauge}) over $z\times \textrm{(3D spacial cylindrical coordinate)}$ in the Euclidian space is
\dis{
\int \, d\theta\, (\frac{dt^2}{2})\, (r^2dr)\, \frac{48\xi a^3}{(r^2+ t^2+a^2)^4} &=4\int \, d\theta\, dt\,  r^2dr\, \left[-\frac{ \xi a^3}{(r^2+ t^2+a^2)^3}\right]_0^\infty =4\int \,    d\theta\,  dr \,  \frac{\xi a^3r^2}{(r^2 +a^2)^3} =\int \,   d\theta\, \frac{ \pi\xi}{4}\,. \label{eq:IntOvThreeD}
}
Thus, from Eq. (\ref{eq:Int1}),
\dis{
 1=\left| \frac{1}{32\pi^2}\int d^4x\, G^\alpha_{\mu\nu}\tilde{G}^{\alpha,\mu\nu} \right| &= \frac{\NDW(a_{MI})}{3! 8\pi^2}\left| \int dz \int d\Omega^z\,  \epsilon_{z\nu\rho\sigma}
\partial^{[\sigma}\,B^{\nu\rho]}\right|\to \int d\theta\, \frac{\NDW(a_{MI})}{3!(2\pi)(4\pi) }\,\frac{ \pi\xi}{4}.\label{eq:Int2}
}
Eq. (\ref{eq:Int2}) is satisfied for the azimuthal angle shift $\theta\to\theta+2\pi$ in the cylindrical coordinate. Thus, $\xi$ is chosen as $ 96/ \NDW(a_{MI})$.  $\NDW(a_{MI})$ is defined as the period  $a_{MI}\to a_{MI}+2\pi\NDW(a_{MI})f_{a_{MI}}$. But, as mentioned before, the mapping $a_{MI}\to\theta$ requires the shift $a_{MI}\to a_{MI}+2\pi f_{a_{MI}}$.  Thus, we determine $\NDW(a_{MI})=1$, and the MI axion domain wall is attached to the string, which is a superstring \citep{WittenMD}, decided by the gauge function $\Lambda_\theta$. When the Universe cools down below the MI axion scale $f_{a_{MI}}$, this class of gauge functions, continuously connected to the above $\Lambda_\theta$, leads to a consistent string-wall system. For other classes of gauge functions, consistent string-wall systems are not derived from superstrings which were originally present above $T> f_{a_{MI}}$.
 
 The axion decay constant $f_{a_{MI}} $ of Eq. (\ref{eq:fvsfp}) is consistent with the domain wall number 1 of the MI axion. The axion mass corresponding to $f_{a_{MI}} $ is marked as the white square in Fig. \ref{fig:Uanom}. The coupling $c_{a\gamma\gamma}$ shown in the vertical axis is the unification point because the MI axion couples to all gauge anomalies with the same coefficient \citep{KimAxCoupl16}. 

\section{$\NDW$ with anomalous U(1)$_{\rm anom}$ below compactification scale}

The MI axion discussed in Sec. \ref{sec:2} applies when it survives as a physical degree below the compactification scale. In the compactification of the heterotic $\EE8'$ string, some U(1) subgroups of $\EE8'$ survive down to low energy. With respect to the surviving low energy fermions, if all these  
U(1) subgroups do not have any gauge anomaly, then the MI axion is a good candidate for the QCD axion with the coupling marked as the white square in Fig. \ref{fig:Uanom}.
   
However, it was pointed out that there arise situations where the compactification process introduces an anomalous U(1)$_{\rm anom}$ gauge symmetry at low energy \citep{AnomUone}. Indeed, explicit models were found to realize such a situation \citep{KimPLB88,Casas89,KimPLB14}. If there arises such an anomalous gauge symmetry U(1)$_{\rm anom}$, it must be a fictitious anomalous U(1).  The way the transverse degrees of U(1)$_{\rm anom}$ are removed is by absorbing the MI axion as its longitudinal degree, $(\partial^\mu) a_{MI}A^{\rm anom}_\mu\equiv (1/M_c) \epsilon^{\mu\nu\rho\sigma} A^{\rm anom}_\mu H_{\nu\rho\sigma}$, and the dynamical degree $a_{MI}$ is removed.\footnote{In gauge theory, the Higgs mechanism gives the gauge boson mass as $\frac{M_A^2}{2}(A_\mu-\partial_\mu a/v)^2 =\frac{M_A^2}{2}(A_\mu)^2+\frac{M_A^2}{2v^2} (\partial_\mu a)^2 -(M_A/v)^2A^\mu \partial_\mu a.$ So, the presence of mixing term removes the kinetic energy term of $a$ and the coefficient in the mixing term is the gauge boson mass.}
The removed U(1)$_{\rm anom}$ gauge boson mass is   $\propto f_{a_{MI}}$ where the coefficient is $\approx (\textrm{compactification scale/string scale})$. Thus, the gauge symmetry U(1)$_{\rm anom}$ is not   present at low energy, and {\it the original superconducting superstring present above $T>f_{a_{MI}}$ loses the domain wall and becomes just superstring without superconductivity} because it lost the domain wall in which the gauge charges flew.
  
Since the U(1)$_{\rm anom}$ gauge boson does not appear at low energy, its effects in the effective interaction of low energy Lagrangian is through the non-renormalizable terms suppressed by the anomalous gauge boson mass $M_A$. Originally the matter fields carried U(1)$_{\rm anom}$ charges, and  they are assigned with the same charges again below the scale $M_A$. These newly defined charges are now the charges of a {\it new} global symmetry U(1)$_\Gamma$ because the gauge boson is removed. The superpotential terms do not involve the space-time derivatives and hence respect the {\it new} global symmetry U(1)$_\Gamma$.  The U(1)$_\Gamma$ is broken by the gauge anomaly and hence can be a good candidate for a PQ symmetry with $f_a$ determined by the VEV of $\sigma$. Then, the cosmological string is expected to be created along the conventional wisdom. In our case, the original superstring which has lost the domain wall becomes the cosmological string with the new domain wall(s) attached to the string by the {\it new} global symmetry U(1)$_\Gamma$, which will be cosmologically realized at the QCD scale.   The gauge anomalies of   U(1)$_\Gamma$ are calculated in \citep{KimPLB88,KimPLB14}.  The interactions mediated by the heavy U(1)$_{\rm anom}$ gauge bosons also respect the   U(1)$_\Gamma$  symmetry since the matter fields have the same charges under U(1)$_{\rm anom}$ and U(1)$_\Gamma$.
Thus, the VEV $\langle\sigma\rangle=f_a/\sqrt2$ gives the green line in the $c_{a\gamma\gamma}$ vs. $m_a$ plane of Fig. \ref{fig:Uanom}, and the cosmologically interesting region around $f_a=10^{10}-10^{11\,}\gev$ is allowed from string compactification.

In models with a PQ-charged singlet field $\sigma$ added in the minimal supersymmetric SM (\SMSSM), the SM fields do not couple to $\sigma$ at the renomalizable level. The lowest order coupling of the SM fields to $\sigma$ is the $d=4$ superpotential term \citep{KimNilles84,KimPRL13},
\dis{
W=\frac{1}{M}H_uH_d\,\sigma^2
}
which is basically the interlocking relation between the global charges of the intermediate scale and the electroweak scale, and its coupling to photon lies on the green line in Fig. \ref{fig:Uanom}.
Now, we  discuss two important cosmological aspects of the axion from U(1)$_\Gamma$ which is called  `$a_\Gamma$' in this paper.

\vskip 0.2cm
\noindent{\it Raising the VEV of $\sigma$}: Let us remind the examples studied in Refs.   \citep{KimPLB88,KimPLB14}. The smallest $\Gamma$ quantum number of \citep{KimPLB88} is 1 and the coefficient of $(g_3^2/32\pi^2)G^\alpha \tilde{G}^\alpha $ is 120. In principle, there can be a $\sigma$ whose $\Gamma$ quantum number is 120. However, there is no such $\Gamma$ in the tables of quarks and doublets of \citep{KimPLB88} as large as 120, and   are at most of order O(10).  Thus, we expect that the VEV $\sigma$ may not be raised as large as by the factor of 120. The singlet charges are not listed, but there will be no $\sigma$ with $\Gamma=120$. 

In another calculation  \citep{KimPLB14} with the model of \citep{Huh09}, many  $\Gamma$ quantum numbers of the matter fields are relatively prime with the coefficient 6984 of $(g_3^2/32\pi^2)G^\alpha \tilde{G}^\alpha $.  In Ref. \citep{KimPLB14} also, the neutral singlets are not listed. However, we can discuss what is expected there, assuming that the $\Gamma$ quantum numbers of neutral singlets behave similarly as the charged singlets. If the $\Gamma$ quantum number $n$ of $\sigma$ is relatively prime with 6984, then the domain wall number is $n$. The largest  $\Gamma$ quantum number among charged singlets, dividing 6984, is --72 \citep{KimAxCoupl16}. If $\sigma$ has $\Gamma=-72$, then  the domain wall number is 97. In any case, there seems to be the domain wall problem. But, the largest possible VEV of $\sigma$ can be as large as  $10^3\times f_{a_{\Gamma}}$ since there are   singlets, carrying such a large $\Gamma$ quantum numbers. In that case,  for $f_{a_{\Gamma}}\approx 10^{11\,}\gev$, the VEV of  $\sigma$ can be  about $10^{14\,}\gev$.  

\vskip 0.2cm
\noindent{\it Domain wall number}: Even  though the VEV of  $\sigma$ can be closer to the GUT scale, the domain wall problem is not solved. It has been known that the physical domain wall number must be modded out by the degeneracy in families unified GUTs \citep{LS82,ChoiKimDW85}. Our case does not belong here. But it is the case envisioned first in Ref. \citep{KimPLB88} where the Goldstone boson direction identifies different vacua. Originally, we had the MI axion $a_{MI}$ and at low energy there is the new axion $a_{\Gamma}$. There are two field directions $a_{MI}$ and $a_{\Gamma}$. 
Their couplings to the gluon anomaly are
\dis{
\frac{1}{32\pi^2}G^\alpha \tilde{G}^\alpha \left(\frac{a_{MI}}{f_{a_{MI}}} +\frac{a_{\Gamma}}{f_{\Gamma}} \right)\label{eq:StringAxCoupl}
}
with the periodicities
\dis{
&a_{MI}\to 2\pi (1)f_{a_{MI}} \\
&a_{\Gamma}\to 2\pi (N_\Gamma)f_{a_{\Gamma}}
}
where their domain wall numbers 1 and $N_\Gamma$ are shown. From Eq. (\ref{eq:StringAxCoupl}), the QCD axion is the combination \citep{ChoiKim85}
\dis{
 \frac{a_{MI}}{f_{a_{MI}}} +\frac{a_{\Gamma}}{f_\Gamma}= \frac{f_\Gamma a_{MI}+f_{a_{MI}} a_\Gamma}{\sqrt{f^2_{a_{MI}}+f^2_\Gamma }} \longrightarrow  a_\Gamma  \textrm{~ in the limit of }   f_{\Gamma}\ll f_{a_{MI}},\label{eq:QCDax}
}
and the Goldstone boson direction is
\dis{
 \frac{f_{a_{MI}} a_{MI} - f_{\Gamma}a_{\Gamma}}{\sqrt{f^2_{a_{MI}}+f^2_\Gamma }}\longrightarrow a_{MI}  \textrm{~ in the limit of }   f_{\Gamma}\ll f_{a_{MI}}.\label{eq:Goldstone}
}
In fact, the longitudinal degree($\approx a_{MI}$) of the anomalous gauge boson is the Goldstone boson direction. The first-glance domain wall number of $a_\Gamma$ seems to be $N_\Gamma$. But, the longitudical direction of the anomalous gauge boson,  $a_{MI}$, identifies all the $N_\Gamma$ vacua of $a_\Gamma$.  Thus, the domain wall number of $a_\Gamma$ is one.

\begin{figure}[!t]
\begin{center}
\includegraphics[width=0.7 \linewidth]{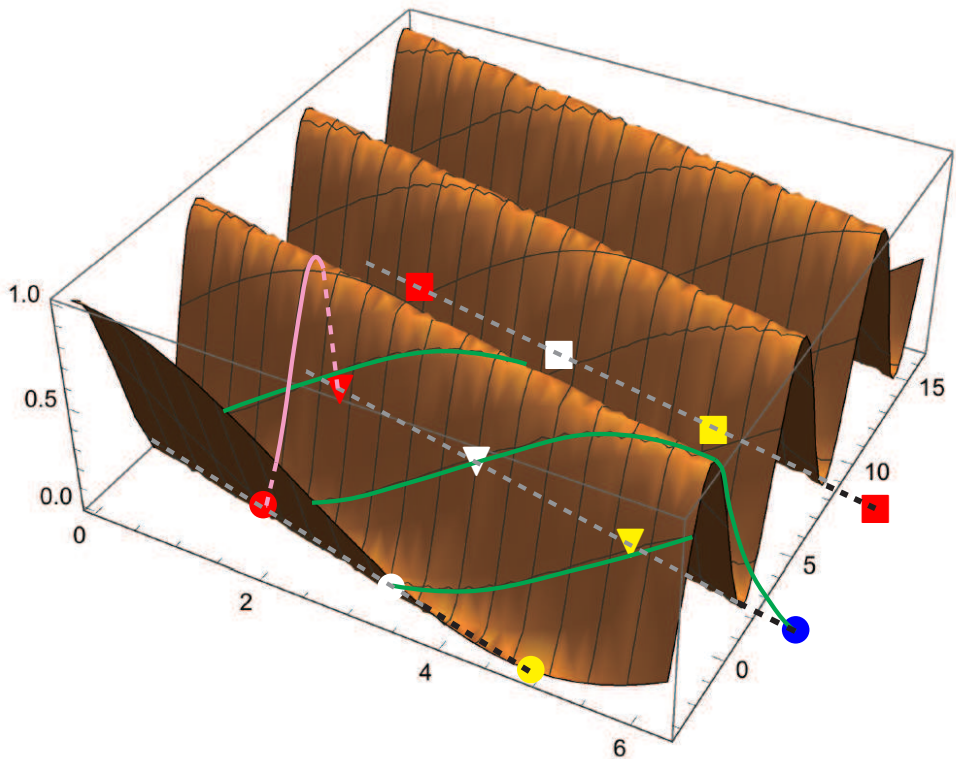} 
\end{center}
\caption{The QCD axion potential from U(1)$_{\rm anom}$. For $n(a_\Gamma)=3$, the vacua are marked with bullets, triangles and squares. The direction of  $a_\Gamma$ is skeched by a lavender curve,  and  the direction of  $a_{MI}$  is skeched by green curves.  The Goldstone boson direction is the flat valley, and all the ridges are the QCD axion potential, and they are the same ones. } \label{fig:DWs}
\end{figure}

 The QCD axion potential from spontaneously broken U(1)$_{\rm anom}$ global symmetry is shown in Fig. \ref{fig:DWs} for $n(a_\Gamma)=3$. Two orthogonal axes of $a_\Gamma$ and $a_{MI}$ are marked as the lavender and green curves, respectively. Three different $a_\Gamma$ vacua are marked with bullets, triangles and squares. For the $a_{MI}$ vacua, the red and blue bullets describe the same vacuum since $\NDW(a_{MI})=1$. In the second valley, the blue bullet is connected to the red triangle by the Goldstone boson (the longitudinal direction of the anomalous U(1) gauge boson) shift, Eq. (\ref{eq:Goldstone}). Thus, one can identify the red bullet and the red triangle, which were originally considered to be different $a_\Gamma$ vacua, are in fact identical.   In this way, one can identify the three vacua. Namely, the first valley is one vacuum which is connected by the Goldstone boson shift. The second valley looks like another valley but it is the same as  first valley, except that there must arise a cosmological domain wall between them.  
 
 Another way to look at Fig. \ref{fig:DWs} is that in the first valley three vacua (red, white, and yellow bullets) are connected by the flat (Goldstone boson) direction, and there is no wall between them. Thus, they are identical.
 
\section{Conclusion}
We have shown that the string compactifications  with an anomalous U(1) gauge symmetry can lead to a cosmologically desirable QCD axion with mass around $10^{-5}-10^{-4\,}\eV$. The VEV of the singlet in the \SMSSM~ can be closer to the GUT scale but the domain wall problem must be resolved within this scheme. In fact,  the longitudinal degree of the anomalous gauge boson identifies the seemingly different vacua and the physical domain wall number of $a_\Gamma$ is $\NDW(a_\Gamma)=1$, allowing  $f_\Gamma\simeq 10^{11\,}\gev$ by the VEV of $\sigma$.
 
\acknowledgments{
This work has been finished at the isolated island Zingst. 
This work is supported in part by the National Research Foundation (NRF) grant funded by the Korean Government (MEST) (NRF-2015R1D1A1A01058449) and  the IBS (IBS-R017-D1-2016-a00).}
\vskip 0.3cm



\begin{thebibliography}{99}
\def\prp#1#2#3{{Phys.\,Rep.}  {\bf #1}  (#3) #2}
\def\rmp#1#2#3{{Rev. Mod. Phys.}  {\bf #1} (#3) #2}
\def\npb#1#2#3{{ Nucl.\,Phys.\,B}   {\bf #1}  (#3) #2}
\def\plb#1#2#3{{Phys.\,Lett.\,B}   {\bf #1}  (#3) #2}
\def\prd#1#2#3{{Phys.\,Rev.\,D}   {\bf #1}  (#3) #2}
\def\prl#1#2#3{{Phys.\,Rev.\,Lett.}   {\bf #1} (#3) #2}
\def\jhep#1#2#3{{JHEP}   {\bf #1} (#3) #2}
\def\jcap#1#2#3{{JCAP}   {\bf #1}  (#3) #2}
\def\zp#1#2#3{{Z.\,Phys.}   {\bf #1} (#3) #2}
\def\njp#1#2#3{{New\,J.\,Phys.}   {\bf #1}  (#3) #2}
\def\epjc#1#2#3{{Euro.\,Phys.\,J.\,C}    {\bf #1} (#3) #2}
\def\frp#1#2#3{{Front.\,Phys.}    {\bf #1}  (#3) #2}
\def\jpg#1#2#3{{J.\,Phys.\,G}   {\bf #1}  (#3) #2}
\def\ijmpd#1#2#3{{Int.\,J.\,Mod.\,Phys.\,D}   {\bf #1} (#3) #2}
\def\mpla#1#2#3{{Mod.\,Phys.\,Lett.\,A}   {\bf #1} (#3) #2}
\def\apj#1#2#3{{Astrophys.\,J.}    {\bf #1}  (#3) #2}
\def\nat#1#2#3{{Nature}    {\bf #1} (#3) #2}
\def\sjnp#1#2#3{{Sov.\,J.\,Nucl.\,Phys.}   {\bf #1} (#3) #2}
\def\apj#1#2#3{{Astrophys.\,J.}   {\bf #1}  (#3) #2}
\def\mnra#1#2#3{{Mon.\,Not.\,Roy.\,Astron.\,Soc.}    {\bf #1} (#3) #2}
\def\jetpl#1#2#3{{JETP\,Lett.}   {\bf #1}  (#3) #2}
\def\pthp#1#2#3{{Prog.\,Theor.\,Phys.}    {\bf #1} (#3) #2}
\def\jkps#1#2#3{{J.\,Korean\,Phys.\,Soc.}   {\bf #1} (#3) #2}
\def\dum#1#2#3{{\bf #1} (#3) #2}

\def\ibid#1#2#3{{\it ibid.}   {\bf #1} (#3) #2}
\def\err#1#2#3{{\bf #1}  (#3) #2\,(E)}   

\bibitem{kim14} For a review, see, J.E. Kim, Y.K. Semertzidis, and S. Tsujikawa, \emph{Bosonic coherent motions in the Universe}, \frp{2}{60}{2014} [arXiv:  1409.2497 [hep-ph]].

\bibitem{PQ77} R. D. Peccei and H. R. Quinn, \emph{CP conservation in the presence of instantons},  \prl{38}{1440}{1977} [doi: 10.1103/PhysRevLett.38.1440].
  
\bibitem{WW78} S. Weinberg, \emph{A new light boson?}, \prl{40}{223}{1978} [doi:: 10.1103/PhysRevLett.40.223];\\
F. Wilczek,  \emph{Problem of strong P and T invariance in the presence of instantons}, \prl{40}{279}{1978} [doi:10.1103/PhysRevLett.40.279].

\bibitem{KSVZ1}  J.E. Kim, \emph{Weak interaction singlet and strong CP invariance}, \prl{43}{103}{1979} [doi: 10.1103/PhysRevLett.43.103].

\bibitem{KSVZ2}  M.A. Shifman, V.I. Vainshtein, V.I. Zakharov, \emph{Can confinement ensure natural CP invariance of strong interactions?}, \npb{166}{493}{1980} [doi:10.1016/0550-3213(80)90209-6].

\bibitem{DFSZ} M. Dine, W. Fischler and M. Srednicki, \emph{A simple solution to the strong CP problem with a harmless axion}, \plb{104}{199}{1981} [doi:10.1016/0370-2693(81)90590-6];\\
 A. P. Zhitnitsky, \emph{On possible suppression of the axion hadron interactions  (in Russian)}, Sov. J. Nucl. Phys. {\bf 31}, 260 (1980), Yad. Fiz. {\bf 31} (1980) 497.


\bibitem{BCMorig83} J. Preskill, M.B. Wise, and F. Wilczek, \emph{Cosmology of the invisible axion}, \plb{120}{127}{1983} [doi:10.1016/0370-2693(83)90637-85]; \\
L.F. Abbott and P. Sikivie, \emph{Cosmological bound on the invisible axion},  \plb{120}{133}{1983} [doi:10.1016/0370-2693(83)90638-X6];\\
M.  Dine and W. Fischler, \emph{The not so harmless axion},  \plb{120}{137}{1983}[doi:10.1016/0370-2693(83)90639-17].  

\bibitem{SikivieDW} P. Sikivie, \emph{Of axions, domain walls and the early Universe}, \prl{48}{1156}{1982} [doi: 10.1103/PhysRevLett.48.1156];\\
A. Vilenkin and A.E. Everett, \emph{Cosmic strings and domain walls in models with Goldstone and pseudoGoldstone bosons}, \prl{48}{1867}{1982} [doi:10.1103/PhysRevLett.48.1867];\\
 S.M. Barr, K. Choi, and J.E. Kim, \emph{Axion cosmology in superstring models}, \npb{283}{591}{1987} [doi:10.1016/0550-3213(87)90288-4];\\
  S.M. Barr and J.E. Kim, \emph{New confining force solution of QCD axion domain wall problem}, \prl{113}{241301}{2014} [arXiv:
1407.4311 [hep-ph]].

\bibitem{KimPRP87} J.E. Kim, \emph{Light pseudoscalars, particle physics, and cosmology}, \prp{150}{1}{1987} 
[doi:10.1016/0370-1573(87)90017-2].

\bibitem{SNstudy88} G.G. Raffelt, \emph{Bounds on exotic particle interactions from SN 1987a}, \prl{60}{1793}{1988} [doi: 10.1103/PhysRevLett.60.1793];\\
M.S. Turner,  \emph{Axions from SN1987a}, \prl{60}{1797}{1988} [doi:  10.1103/PhysRevLett.60.1797]\\
K. Choi, K. Kang, and J.E. Kim, \emph{Invisible axion emissions from SN1987A}, \prl{62}{849}{1989} [doi: 10.1103/PhysRevLett.62.849].

\bibitem{GGW81}  M.B. Wise, H. Georgi, and S.L. Glashow,   \emph{SU(5) and the invisible axion}, \prl{47}{402}{1981} [doi:10.1016/0550-3213(81)90433-8].

\bibitem{GS84} M.B. Green and J. Schwarz, \emph{Anomaly cancellation in supersymmetric D=10 gauge theory and superstring theory}, \plb{149}{117}{1984} [doi:10.1016/0370-2693(84)91565-X].
 
\bibitem{WittenMI} E. Witten, \emph{Some properties of O(32) superstrings}, \plb{149}{351}{1984} [doi:10.1016/0370-2693(84)90422-2].

\bibitem{WittenMD} E. Witten, \emph{Cosmic superstrings}, \plb{153}{243}{1985} [doi:10.1016/0370-2693(85)90540-4].

\bibitem{Nilles84} H.P. Nilles, \emph{Supersymmetry, supergravity and particle physics}, \prp{110}{1}{1984} [doi:10.1016/0370-1573(84)90008-5].
 
\bibitem{ChoiKim85} K. Choi and J. E. Kim, \emph{Harmful axions in superstring models}, \plb{154}{393}{1985} and \err{156}{452}{1985} [doi:10.1016/0370-2693(85)90416-2].

\bibitem{KimPLB88} J. E. Kim, \emph{The strong  CP problem in orbifold compactifications and an SU(3)$\times$SU(2) $\times$U(1)$^n$ model}, \plb{207}{434}{1988} [doi:10.1016/0370-2693(88)90678-8].

\bibitem{KimPLB14} J. E. Kim, \emph{Calculation of axion--photon--photon coupling in string theory}, \plb{735}{95}{2014}  and \err{741}{327}{2014} [arXiv:1405.6175 [hep-ph]]. 

\bibitem{LS82}   G. Lazarides and Q. Shafi, \emph{Axion models with no domain wall problem}, \plb{115}{21}{1982} [doi:10.1016/0370-2693(82)90506-8]. 

\bibitem{ChoiKimDW85}  K. Choi and J.E. Kim, \emph{Domain walls in superstring models}, \prl{55}{2637}{1985}  [doi: 10.1103/PhysRevLett.55.2637].

\bibitem{BarrGr92}
S. M. Barr and D. Seckel, \emph{Planck scale corrections to axion models}, \prd{46}{539}{1992} [doi: 10.1103/PhysRevD.46.539];\\
M. Kamionkowski and J. March-Russell,
\emph{Planck scale physics and the Peccei-Quinn mechanism}, \plb{282}{137}{1992} [hep-th/9202003];\\
R. Holman, S. D. H. Hsu, T. W. Kephart, E. W. Kolb, R. Watkins, and L. M. Widrow, \emph{Solutions to the strong CP problem in a world with gravity}, \plb{282}{132}{1992} [hep-ph/9203206];\\
B. A. Dobrescu, \emph{The strong CP problem versus Planck scale physics}, \prd{55}{5826}{1997} [hep-ph/9609221].

\bibitem{KimPLB13} J.E. Kim, \emph{Abelian discrete symmetries $\Z_N$ and $\Z_{nR}$ from string orbifolds}, \plb{726}{450}{2013} [arXiv: 1308.0344[hep-th]].
  
\bibitem{ChoiKimIW07}  K.-S. Choi, I.-W. Kim and J. E. Kim, \emph{String compactification, QCD axion and axion-photon-photon coupling},  \jhep{0703}{116}{2007} [arXiv:hep-ph/0612107].

\bibitem{KimAxCoupl16} J.E. Kim, arXiv:1603.02145 [hep-ph].

\bibitem{KimPRD98} J.E. Kim, \emph{Constraints on very light axions from cavity experiments}, \prd{58}{055006}{1998} [arXiv: hep-ph/9802220 ]. 

\bibitem{KimRMP10} J. E. Kim and G. Carosi, \emph{Axions and the strong CP problem}, \rmp{82}{557}{2010} [arXiv: 0807.3125[hep-ph]].

\bibitem{Belavin74}  A.A. Belavin, A. Polyakov, A. Schwartz, and Y. Tyupkin, \emph{Pseudoparticle solutions of the Yang-Mills equations}, \plb{59}{85}{1974} [doi:  10.1016/0370-2693(75)90163-X].

\bibitem{Kibble80prp} T.\,W.\,B. Kibble, \emph{Some implications of a cosmological phase transition}, \prp{67}{183-199}{1980} [doi: 10.1016/0370-1573(80)90091-5]. 

\bibitem{AnomUone} J.J. Atick, L. Dixon, and A. Sen, \emph{String calculation of Fayet-Iliopoulos $d$ terms in arbitrary supersymmetric compactifications}, \npb{292}{109}{1987} [doi:10.1016/0550-3213(87)90639-0];\\
M. Dine, I. Ichinose, and N. Seiberg, \emph{F terms and $d$ terms in string theory}, \npb{293}{253}{1987} [doi:10.1016/0550-3213(87)90072-1].

\bibitem{Casas89} A. Casas, E.K. Katehou, and C. Munoz, \emph{U(1) Charges in orbifolds: Anomaly cancellation and phenomenological consequences}, \npb{317}{171}{1989} [doi: 10.1016/0550-3213(89)90566-X].

\bibitem{KimNilles84} J.E. Kim and H.P. Nilles, \emph{The $\mu$ problem and the strong CP problem}, \plb{138}{150}{1984} [doi:10.1016/0370-2693(84)91890-2].
  
\bibitem{KimPRL13} J.E. Kim, \emph{Natural Higgs-flavor-democracy solution of the $\mu$ problem of supersymmetry and the QCD axion}, \prl{111}{031801}{2013} [arXiv:1303.1822 [hep-ph]].

\bibitem{Huh09} J-H Huh, J.E. Kim, and B. Kyae, \emph{SU(5)$_{\rm flip}\times$SU(5)$'$}, \prd{80}{115012}{2009}  [arXiv: 0904.1108 [hep-ph]].

\end{thebibliography}
\end{document}